\documentclass[twocolumn,aps,nofootinbib,superscriptaddress]{revtex4-1}

\usepackage{graphicx}
\usepackage{booktabs}
\usepackage{xcolor}
\usepackage{hyperref}
\usepackage{multirow}
\usepackage{setspace}
\usepackage{amsmath, amssymb, amsthm}
\usepackage{cleveref}
\usepackage{physics}
\usepackage{soul}

\DeclareMathOperator*{\argmax}{argmax} 
\DeclareMathOperator{\median}{median}
\DeclareMathOperator{\sign}{sign}

\usepackage{times}
\newcommand{\etal}{\textit{et al.}}

\newcommand{\beginsupplement}{%
        \setcounter{table}{0}
        \renewcommand{\thetable}{S\arabic{table}}%
        \setcounter{figure}{0}
        \renewcommand{\thefigure}{S\arabic{figure}}%
        \renewcommand{\thesection}{S\arabic{section}}
     }

\begin{document}
\title{Negative Representation and Instability in Democratic Elections}
\author{Alexander F. Siegenfeld} 
\email{Corresponding author.  Email: asiegenf@mit.edu}
\affiliation{Department of Physics, Massachusetts Institute of Technology, 77 Massachusetts Ave., Cambridge, MA}
\affiliation{New England Complex Systems Institute, 277 Broadway, Cambridge, MA}
\author{Yaneer Bar-Yam}
\affiliation{New England Complex Systems Institute, 277 Broadway, Cambridge, MA}

\maketitle

\textbf{The challenge of understanding the collective behaviors of social systems can benefit from methods and concepts from physics~\cite{Complex2018,savit1999adaptive,stauffer2008social,castellano2009statistical,kitsak2010identification,ratkiewicz2010characterizing}, not because humans are similar to electrons, but because certain large-scale behaviors can be understood without an understanding of the small-scale details~\cite{bar2016big}, in much the same way that sound waves can be understood without an understanding of atoms.
Democratic elections are one such behavior.   Over the past few decades, physicists have explored scaling patterns in voting and the dynamics of political opinion formation, e.g.~\cite{fortunato2007scaling,galam2008sociophysics,borghesi2010spatial,chatterjee2013universality,fernandez2014voter,braha2017voting}.  
Here, we define the concepts of \textit{negative representation}, in which a shift in electorate opinions produces a shift in the election outcome in the opposite direction, and \textit{electoral instability}, in which an arbitrarily small change in electorate opinions can dramatically swing the election outcome, and prove that unstable elections necessarily contain negatively represented opinions.  
Furthermore, in the presence of low voter turnout, increasing polarization of the electorate can drive elections through a transition from a stable to an unstable regime, analogous to the phase transition by which some materials become ferromagnetic below their critical temperatures.  
Empirical data suggest that United States presidential elections underwent such a phase transition in the 1970s and have since become increasingly unstable.}  

Elections are, fundamentally, a means of aggregating many opinions into one---those of the citizens into that of the elected official.  Here, an opinion refers to a person's entire set of political beliefs; i.e. each citizen (and candidate) has one opinion.  For simplicity, we focus on the case in which the set of all possible opinions can be embedded in a one-dimensional continuous space (e.g. a position on a left-right spectrum), as in many studies in the social choice literature (reviewed in \cref{screview}).  However, our results can be extended to a multidimensional space (see \cref{multi}).
The first part of this manuscript examines general properties of electoral representation (which we connect to the voting power literature in \cref{consistency}) and instability, using a mathematical formalism that departs from the literature in that it makes no assumptions about how people vote or even the structure of the voting process.  The second part of this manuscript presents a specific model that builds upon the social choice literature in order to demonstrate how, when the assumption of concave voter preferences is relaxed, instability and negative representation can arise.  The specific model is shown to map onto the well-known mean-field Ising model~\cite{Kadanoff2009} of magnetic materials.  The emergence of instability can couple to geospatially varying local election outcomes and the potentially destabilizing effects of a two party system.  Finally, the manuscript considers the implications of such a model for contemporary American elections.

We define an \textit{election} by $y[f(x)]$, a \textit{functional} that maps the distribution of electorate opinions $f(x)$---defined so that for any interval $[a,b]\subset\mathbb{R}$, the number of citizens with opinions in that interval is $\int_a^bf(x)dx$---to the election outcome $y\in\mathbb{R}$, i.e. the opinion of the elected official.  Note that in this framework, candidacy is endogenous: candidate opinions---or equivalently, which candidates run---are themselves functions of the electorate opinions.  Any electoral system, regardless of its detailed mechanisms (e.g. the number of candidates or parties, the existence of primary elections, restrictions on candidate entry, etc.) can be conceptualized as such a process that outputs the opinion of the winning candidate based on the electorate opinions.  In order that no opinion be \textit{a priori} privileged over others, the one restriction we place on this process is \textit{translational invariance}, i.e. 
\begin{equation} 
\label{invar}
y[f(x+c)]+c=y[f(x)]
\end{equation}
for all $c$ (\cref{transinvar}).

     In order to measure the sensitivity of the election outcome to changes in electorate opinions~\cite{Napel2004}, we define the \textit{representation} of an opinion $x$ by
    \begin{equation}
    \label{genrep}
    r_c(f,x)=\frac{\delta y}{c}
    \end{equation} 
where $\delta y$ is the change in outcome that occurs if an individual's opinion changes from $x$ to $x'=x+c$.  (Representation should \textit{not} be defined using only the distance between a citizen's opinion and that of the elected candidate: opinions without causal influence on the election outcome are not represented, even if they happen to align with that outcome.)
It is convenient to measure representation by
    \begin{equation}
    \label{limrep}
    r(f,x)=\lim_{c\rightarrow 0}r_c(f,x)
    \end{equation}
when the limit exists, since $r(f,x)$ does not depend on $c$.

For a large population, a number of results hold if the election is differentiable (see \cref{repproofs} for details).  First,
\begin{equation}
\label{rtilde}
r(f,x)=\frac{d}{dx}\frac{\delta y}{\delta f(x)}
\end{equation}
Second, $r_c(f,x)$ is the average of $r(f,x)$ over the interval $[x,x+c]$, and thus representation of individual opinions can be measured by $r(f,x)$ alone.   
Third, the total representation of the electorate's opinions equals $1$:
\begin{equation}
\label{sumtoone}
\int_{-\infty}^{\infty} f(x) r(f,x)dx=1  
\end{equation}

\begin{figure}
\centering
\includegraphics[width=.5\textwidth]{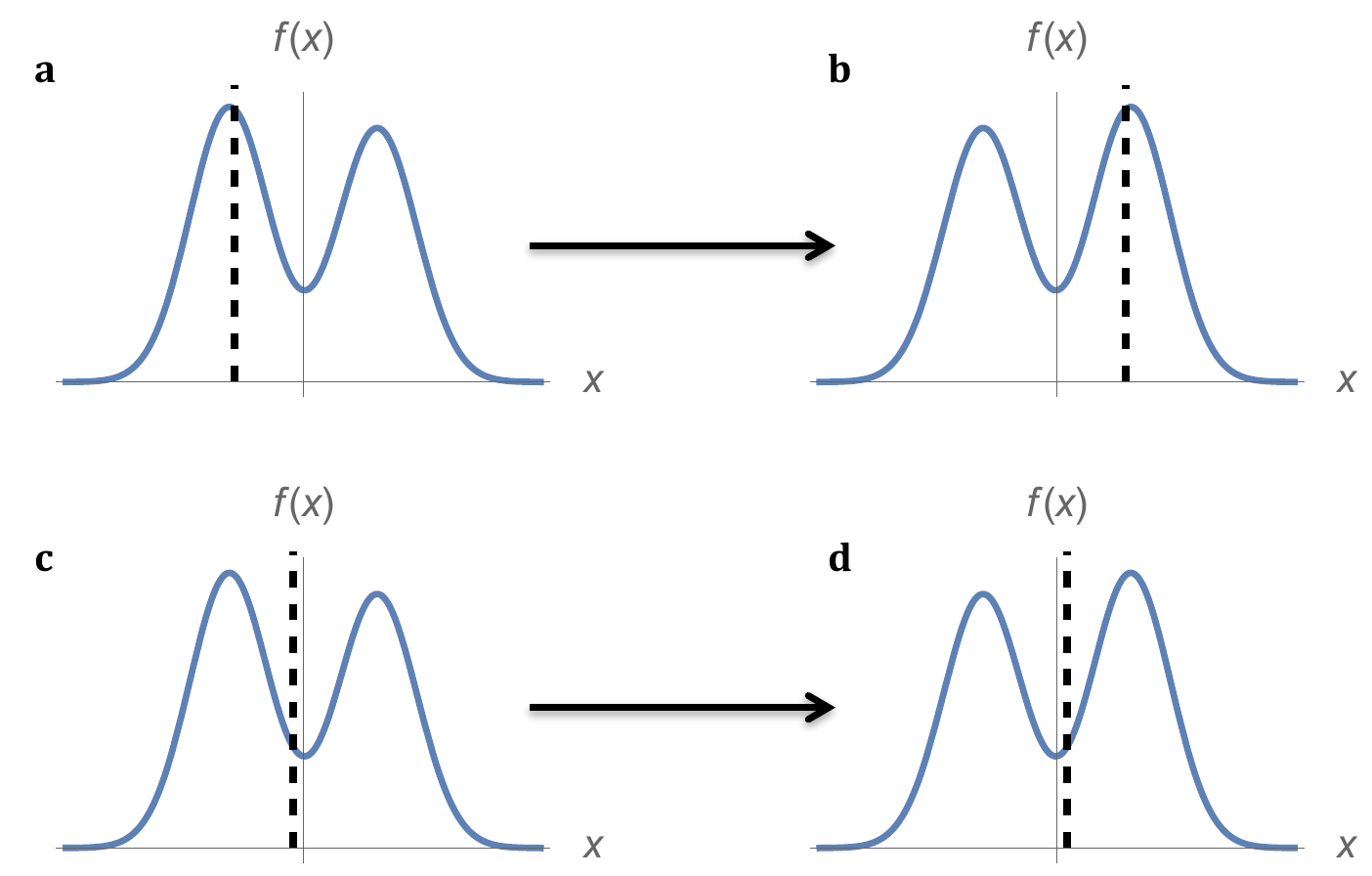}
\caption{Election outcomes (vertical dashed lines) shift in response to a shift in the distribution of electorate opinions ($f(x)$, denoted by the solid curves, where the horizontal axes ($x$) denote political opinion).  \textbf{a-b}, When not everyone votes, an election can be unstable (\cref{instaboutcome})---a small shift in opinions to the right causes a large swing in the election outcome.  \textbf{c-d}, In a stable election, by contrast, a small shift in opinions causes a similarly small shift in outcome.}
\label{instabvsmedian}
\end{figure}

We now show that all unstable elections contain negatively represented opinions.
An election is \textit{unstable} if an arbitrarily small change in opinion can cause a sizable change in the election outcome (\cref{instabvsmedian}), i.e. for some $f$ and $x$, 
\begin{equation}
\label{instab}
\lim_{c\rightarrow 0} c r_c(f,x)\neq 0  
\end{equation}
If an election $y[f]$ is unstable for an opinion distribution $f_0$, then if some opinion $x_0$ changes by a small amount $\epsilon$ (call the resulting opinion distribution $f_1$), the election outcome changes by a larger amount $C$, i.e. $\delta y_1\equiv y[f_1]-y[f_0]=C$ with $|C|>|\epsilon|$.   Now consider starting with $f_0$ and shifting all opinions except $x_0$ by $-\epsilon$ (call the resulting opinion distribution $f_2$).  Since $f_2(x)=f_1(x+\epsilon)$, $\delta y_2\equiv y[f_2]-y[f_0]=C-\epsilon$ by translational invariance (\cref{invar}).  Thus, $\delta y_1$ and $\delta y_2$ have the same sign (since $|C|>|\epsilon|$), despite being caused by changes of opinion in opposite directions, so one of the two changes in opinions must be negatively represented.  
This proof that unstable elections always contain negatively represented opinions relies only on translational invariance; therefore, we expect it to hold generally in real-world conditions, regardless of the size of the electorate, the number of candidates or parties, the existence of primaries, the effects of the Electoral College, etc.  

Thus far we have presented general results that apply to all elections; we now examine a particular class of models in order to illustrate how negative representation can arise and to show how polarization drives elections through a phase transition into the unstable regime.
Consider an election with two candidates (or two major political parties) who choose their positions in order to maximize their chances of winning the election (or, equivalently, choose to run/are selected to run in the general election based on these considerations). 
Then, under the assumptions of an affine linear \textit{utility difference model}~\cite{Banks2005} (see \cref{udmexamples}), the possible election outcomes are given by (Theorem~4 of~\cite{Banks2005}): 
\begin{equation}
 \label{ueqm}
\argmax_{y\in\mathbb{R}} \int_{-\infty}^{\infty}  u_x(y)f(x)dx
 \end{equation}
where the utility function $u_x$ denotes the political preferences of someone with opinion $x$.  We note that this model always admits at least one Nash equilibrium in candidate strategies (\cref{nash}); the instability we describe, which arises from multiple Nash equilibria, should be distinguished from scenarios in which no Nash equilibria exist (\cref{screview}).  Although this model may not accurately describe individual and candidate behaviors, 
it can nonetheless be used to describe the properties of real-world elections, since representation and instability depend on electoral mechanisms only through the effects of these mechanisms on $y[f]$---the functional that characterizes how the election outcome varies with electorate opinions.  For instance, this model cannot describe the deterministic voting underlying the Median Voter Theorem~\cite{Black1948} but can nonetheless exactly capture the collective behavior to which such voting gives rise, namely the election of the median opinion (\cref{udmexamples}).

\begin{figure}
\centering
\includegraphics[width=.5\textwidth]{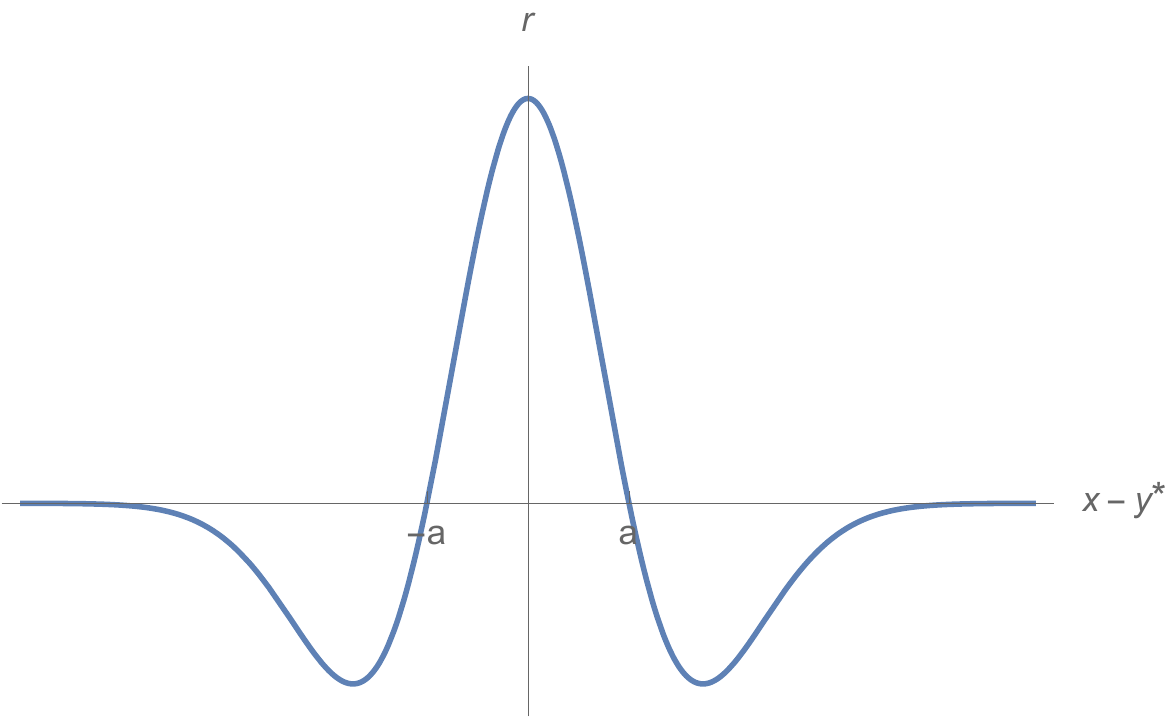}
\caption{When some voters abstain, opinions far from the election outcome may be negatively represented.  This graph depicts the representation ($r$) of opinions ($x$) as a function of their distance from the election outcome ($y^*$) for voting behavior given by \cref{ualien}.}
\label{negrep}
\end{figure}

If citizens with opinions that are far from both candidates are more likely to abstain from voting, a phenomenon known as \textit{alienation}~\cite{hinich1972,hinich1978,Southwell2003,plane2004candidates,Adams2006}, then negative representation occurs.  Since not voting for either candidate is equivalent to preferring them equally, a utility function $u_x(y)$ that captures this behavior will be almost flat for large $|y-x|$.  One such utility function is 
\begin{equation}
u_x(y)=u(y-x)=e^{-\frac{(y-x)^2}{2a^2}}
\label{ualien}
\end{equation}
where $a$ is a positive constant.  Representation is then given by (\cref{udmrep})
\begin{equation}
\label{ralien}
r(f,x)\propto-\frac{1}{N}u''(y^*-x)=\frac{1}{Na^4}(a^2-(y^*-x)^2)e^{-\frac{(y^*-x)^2}{2a^2}}
\end{equation}
where $N$ is the number of constituents.  Opinions far from the election outcome ($|x-y^*|>a$) are negatively represented ($r(f,x)<0$, see \cref{negrep}): the election outcome is inversely sensitive to changes in those opinions.  
For instance, given a center-left candidate and a candidate to the right, as left-wing individuals move farther left, they may become less likely to vote for the center-left candidate (choosing instead not to vote), which increases the probability that the candidate on the right will win.  In response, the electoral equilibrium of future elections may shift rightward, as candidates no longer vie for these left-wing votes.  
(Eq.~\ref{ualien} is just an example; more generally, negative representation will occur if and only if $u(y-x)$ is not concave.)
Thus, individual choices to abstain when neither candidate is appealing lead to a system-level perversion in the aggregation of electorate opinions, in which the electorate becoming more left-wing can result in a more right-wing outcome, or vice versa.

\begin{figure}
\centering
\includegraphics[width=.5\textwidth]{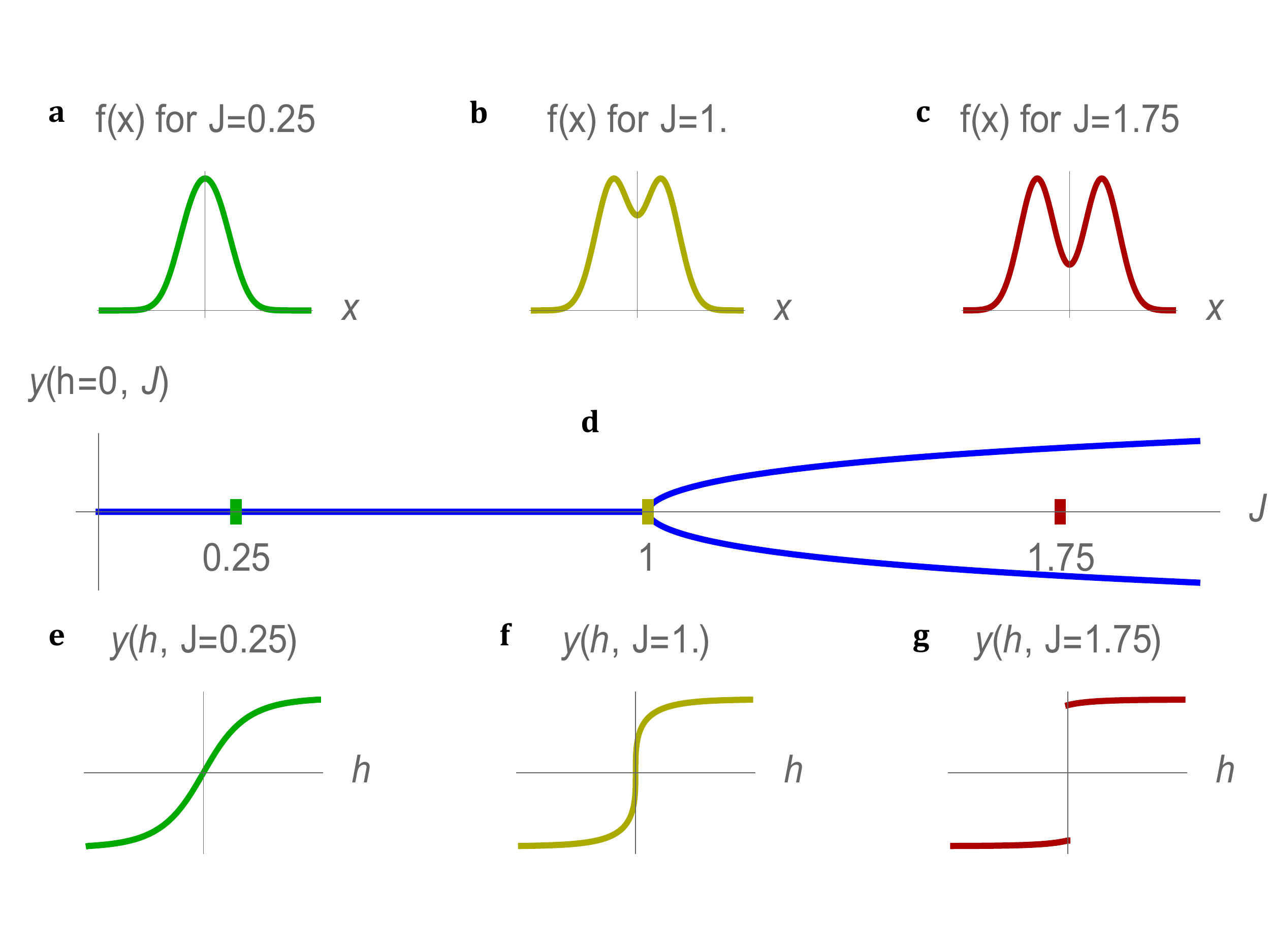}
\caption{The stability of elections depends on the degree of electorate polarization.  \textbf{a-c}, With increasing polarization ($J$) of the electorate opinion distribution, \textbf{d}, the electoral system undergoes a phase transition from possessing a single stable outcome to possessing two possible unstable outcomes.  \textbf{e}, In the stable regime, the outcome smoothly responds to changes in the relative sizes of the two subpopulations (changes in $h$).  \textbf{f}, At the phase transition ($J=1$), the outcome is continuous but not differentiable in the relative sizes of the two subpopulations.  \textbf{g}, In the unstable regime, the outcome discontinuously jumps.  These figures were created using \cref{sumgauss,instaboutcome} for $a=\sigma=1$ (for which $J=\Delta^2/2$).}
\label{bifurcation}
\end{figure}

When combined with a sufficiently polarized electorate, nonvoting (in particular, non-concave $u$) leads not only to negative representation but also to instability. (As proven earlier, instability cannot occur without negative representation; however, negative representation is possible without instability---see \cref{instabproofs} for further discussion.)  For $u_x$ defined by \cref{ualien} and an opinion distribution $f(x)$ consisting of two normally distributed (potentially unequally sized) subpopulations centered at $\pm\Delta$ (without loss of generality, we define the origin as the (unweighted) average of the means of the subpopulations), 
\begin{equation}
\label{sumgauss}
f(x)=w_1 e^{-\frac{(x+\Delta)^2}{2\sigma^2}}+w_2 e^{-\frac{(x-\Delta)^2}{2\sigma^2}},
\end{equation}
the outcome $y$ is given by the following condition:
\begin{equation}
y/\Delta
=\tanh(J y/\Delta+h)
\label{instaboutcome}
\end{equation}
where $J$, a dimensionless measure of the polarization of the electorate, is given by $J=\Delta^2/(a^2+\sigma^2)$ and $h$, a measure of the relative sizes of the two subpopulations of the electorate, is given by $h=\frac{1}{2}\ln\frac{w_2}{w_1}$.  For $w_1=w_2$ (i.e. for equally sized subpopulations), $h=0$, and the election is stable for $J\leq1$ and unstable for $J>1$.  In the stable regime, $y=0$.  In the unstable regime, there are two possible outcomes described by $\pm|y^*|$, and an arbitrarily small change in $h$ can cause $y$ to swing between its positive and negative values (\cref{bifurcation}).  Thus, small variations determine which subpopulation wins. The variations might include changes in population, nuances in the candidates' personalities, changes in the rules (simple majority versus Electoral College system, for example), voting restrictions, and the effectiveness of turnout operations.  From election cycle to election cycle, the outcome can swing between the two subpopulations, with the majority of the opinions in the losing subpopulation being negatively represented (\cref{instabproofs}).  An intuitive explanation of how such negative representation and instability arises is that, due to low voter turnout, candidates are incentivized to focus on turning out their bases rather than winning over centrist voters. 

This voting model (\cref{instaboutcome}) is precisely equivalent to a mean-field Ising model of a ferromagnet~\cite{Kadanoff2009}, in which each spin (magnetic dipole) interacts with every other spin, highlighting the hidden dependencies that can arise from a system in which citizens' behaviors are superficially independent (\cref{ising}).  This connection with the Ising model differs from other applications of the Ising model to complex systems~\cite{baryam1997,michard2005theory,grabowski2006ising,sornette2014physics} in two respects: first, our analysis takes the electorate opinions as given and therefore concerns the instability in the election dynamics rather than in the dynamics of the electorate opinions themselves, and second, we do not impose any of the assumptions of the Ising model, but rather show (\cref{ising}) that such an equivalence naturally arises for certain classes of voting behavior.  Because the citizens are effectively coupled through collectively choosing a candidate, the system exhibits emergent behavior in which there can be discontinuities in the election outcome despite the continuous voting behaviors of individuals.   Such emergent discontinuities are a common feature of complex systems~\cite{baryam1997,May2008,Scheffer2009,Mora2011,Bouchaud2013,harmon2015anticipating}.

These emergent discontinuities can couple to geospatial pattern formation.  Given the existence of local elections, we can consider a spatially varying election outcome $y(\vec{r}_i)$, which serves as an order parameter for the $2D$ geographic system ($\vec{r}_i$ represents geographic location).  Given local interactions, the $2D$ Ising model exhibits a universal behavior of spatial pattern formation~\cite{bray2002theory}.  Indeed, this is consistent with the observed geographic segregation associated with political polarization~\cite{Pew2016}.  (As these dynamics are consistent with social influence~\cite{berelson1954voting} and homophily~\cite{mcpherson2001birds}, we expect the order parameter $y(\vec{r}_i)$ to couple to these forces.)  One such model that exhibits this universal behavior can be constructed by coarse-graining the local electoral outcomes $y(\vec{r}_i)$ into a statistical field~\cite{borghesi2010spatial,borghesi2012election} $\tilde{y}(\vec{r})$ with an effective local hamiltonian (normalized by temperature) that allows for geographic heterogeneity:
\begin{equation}
\int d^2\vec{r}  \qty[t(\vec{r}) \tilde{y}(\vec{r})^2  + u(\vec{r})\tilde{y}(\vec{r})^4 + K(\vec{r}) (\nabla \tilde{y}(\vec{r}))^2 - h(\vec{r})\tilde{y}(\vec{r})]
\end{equation}
 where $u,K:\mathbb{R}^2\rightarrow(0,\infty)$ and $t,h:\mathbb{R}^2\rightarrow \mathbb{R}$.

We can also consider a model in which the political parties are explicit.  Generically, the memberships of the two parties can be expected to differ in opinion due to the forces described in the previous paragraph.  
A model that captures this breaking of symmetry between the two parties can be described by the following temperature-normalized hamiltonian for $p:\mathbb{R}\rightarrow [0,1]$:
\begin{equation}
-\max_{y_1,y_2\in\mathbb{R}}\int_{-\infty}^\infty \beta f(x)\qty[p(x)u_x(y_1)+(1-p(x))u_x(y_2)]dx
\end{equation}
where $p(x)$ denotes the probability that a citizen with opinion $x$ belongs to or leans towards one of the two parties (with $1-p(x)$ being the probability that they align with the other party), where $\beta$ is a positive constant, and where the rest of the notation follows that of \cref{ueqm}.
As long as at least one of these parties chooses its nominee so as to maximize the probability of winning the general election (e.g. if primary voters vote based on this consideration), then no changes need be made to the analysis in \cref{ueqm,ualien,ralien,sumgauss,instaboutcome}.  
But, if \textit{both} parties choose their nominees primarily based upon the opinions of their members rather than upon electability, then differences in the parties' members will generally cause the election to be unstable, regardless of general election voting behavior.  For example, if each of the subpopulations in \cref{sumgauss} approximately corresponds to a political party, then nominees based on the median or mean opinions of the party members will be located at $\pm\Delta$, thus rendering the election unstable for all $\Delta >0$ if the two parties have approximately equal support.  Negative representation must then also be present; one such example would be if some opinions of the left-leaning party were to shift further to the left, resulting in a more left-leaning nominee that is less appealing to general election voters, increasing the likelihood that the right-leaning party's nominee wins.  
Given this inherently polarizing effect of the two-party system, electoral reforms such as instant-runoff voting or approval voting that allow for third-party candidates to run without playing spoiler may reduce this instability.

\begin{figure}
\centering
\includegraphics[width=.5\textwidth]{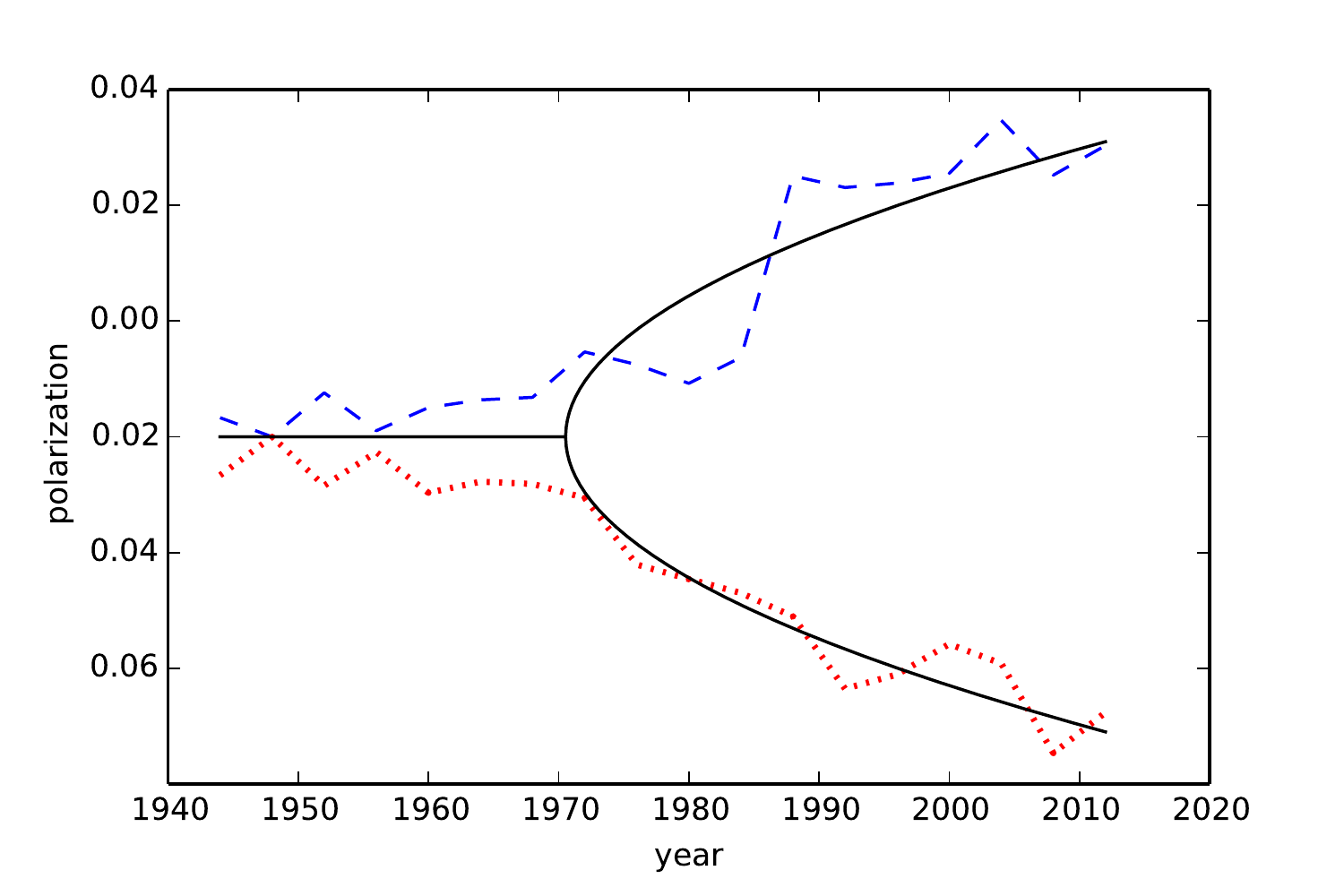}
\caption{The polarization of the Democratic (dashed) and Republican (dotted) parties between 1944 and 2012, as measured by the fraction of polarizing words in the party platforms relative to a baseline.  Party platforms are released once every four years.  During the 1970s, there appears to be a divergence, which may correspond to a phase transition of the electoral dynamics into instability.   The bifurcation associated with the universality class to which the mean-field Ising model belongs (solid, cf. \cref{bifurcation}d) is superimposed; $R^2$ values are $0.86$ for the Democratic party and $0.89$ for the Republican party.  See \cref{empirics} for further details.}
\label{empirical}
\end{figure}

Empirically, the opinions of the United States population, and in particular the opinions of those most likely to vote, have been polarizing over the past few decades~\cite{ref:pew}, while voter turnout has remained approximately constant~\cite{voterturnout}.  Thus, we might expect that over time, election outcomes have undergone a phase transition from a stable to unstable regime, and indeed this appears to be the case (\cref{empirical}).  (While elections cannot be expected to follow the precise assumptions behind \cref{instaboutcome}, they nonetheless may fall into the same \textit{universality class} (\cref{empirics}), including in the case where the space of opinions is multidimensional (\cref{multi}).)  Changes to the electoral process, such as reforms in the early 1970s as to how presidential party nominees are chosen, together with increasing polarization, may have driven this phase transition into instability.
That the policies of legislators in politically homogeneous districts are more strongly correlated with the preferences of the district's median voter than the policies of legislators in heterogeneous districts~\cite{Gerber2004} lends further empirical support.  

In addition to whatever other problems may arise from instability, unstable elections also necessarily contain negatively represented opinions, a result that, due to the generality of the assumptions used to prove it, is expected to hold for any real-world election, regardless of the structure of the electoral process, the number of candidates, or other details.  Therefore, the impact of electoral reforms on this instability should be considered.  For instance, when voter turnout is low, political polarization can fundamentally shift electoral dynamics, causing large swings from election to election and leaving many negatively represented, regardless of the outcome.  These results suggest that polices that increase voter turnout will not only result in more voices being heard but will also stabilize elections and reduce negative representation.  


\subsection*{Acknowledgements}
\vspace{-1em}
This material is based upon work supported by the National Science Foundation Graduate Research Fellowship under Grant No. 1122374 and the Hertz Foundation.  We thank B. Dan Wood for sharing the data from his paper on the polarization of party platforms~\cite{Jordan2014} and Irving Epstein and Mehran Kardar for helpful feedback.

\beginsupplement

\section{Methods}
 
\subsection{Translational invariance}
\label{transinvar}
In the main text, we make the assumption of translational invariance (\cref{invar}).  Technically, translational invariance is defined only in relation to a particular metric.  Thus, the assumption of translational invariance can be relaxed without invalidating the paper's results.  For the proof that negative representation implies instability, all that is required is that the election be continuous (rather than invariant) under translations or formally, that for $f_c(x)=f(x+c)$, $y[f_c]$ be continuous in $c$ (note that this property is independent of the metric).  The proof given in the main text then follows in the limit $\epsilon\rightarrow0$.   For the proof that the total representation sums to $1$ (\cref{sumtoone}), there need only exist some metric on the opinion space under which the election is translationally invariant, so long as the representation is defined under such a metric.

\subsection{Representation in the large-population limit} 
\label{repproofs}

\begin{figure}
\centering
\vspace{2em}
\includegraphics[width=.5\textwidth]{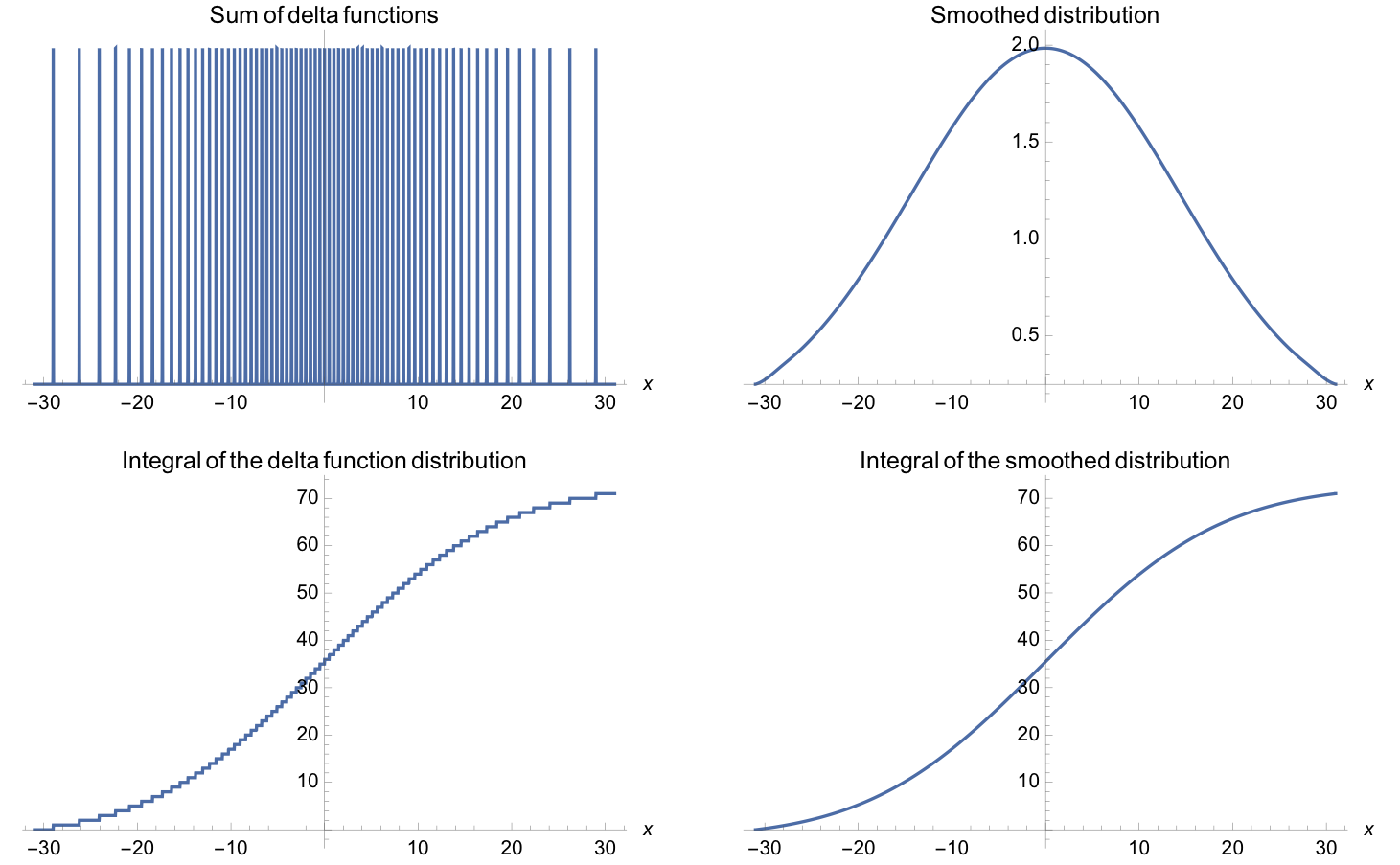}
\caption{By replacing the Dirac delta functions with the approximation $\delta(x)\sim \frac{1}{\sqrt{2\pi}} e^{-x^2/2}$, a smooth distribution is obtained from the sum of delta functions.  Note that the integrals of these distributions (taken from the lower bound of the domain of the graphs to $x$) are very similar, despite the striking differences between the distributions themselves.}
\label{fig:smoothing}
\end{figure}

In this section, we derive properties of our representation measure when the number of citizens $N$ is large.  For a set of $N$ citizens with opinions $\{x_1,x_2,...,x_N\}$, the distribution of electorate opinions is $f(x)=\sum_{i=1}^N\delta(x-x_i)$, which is the only distribution satisfying the property that $\int_a^bf(x)dx$ is the number of citizens with opinions in the interval $[a,b]$.  However, it is often useful to choose $f(x)$ to be a smooth function that approximates $\sum_{i=1}^N\delta(x-x_i)$, in the sense that the difference between $\int_a^bf(x)dx$ and the number of citizens with opinions in the interval $[a,b]$ is no greater than $1$ for all $a,b$ (see \cref{fig:smoothing} for an example).   For large enough $N$, the error of up to $1$ opinion will generally not be significant.  Alternatively, a natural interpretation of a smooth $f(x)$ is that the opinions are themselves probabilistic (for an example of explicitly probabilistic opinions, see \cref{consistency}).  Whether or not $f(x)$ is chosen to be smooth does not matter for the results of the text, although for the results that rely on the assumption that the number of citizens is large, the mathematics are simpler if $f(x)$ is assumed to be a function rather than a distribution.  For instance, the expression for representation in the case of median voting involves evaluating $f$ at its median, an operation which is not well-defined if $f$ is a sum of Dirac delta functions.

In the limit of a large population ($N>>1$), the change $\delta f$ in the opinion distribution arising from an individual opinion will be small compared to the opinion distribution as a whole, and so we expand $\delta y$ to first order in $\delta f$:
\begin{equation}
\label{largen2}
\delta y=\int_{-\infty}^{\infty} \delta f(z)\frac{\delta y}{\delta f(z)}dz
\end{equation}
Note that \cref{largen2} does not apply to cases in which $y[f]$ is not differentiable, e.g. when the election is unstable and small changes in the opinion distribution can have an outsized impact; thus we do not use results derived from these equations when analyzing instability.   

We now derive \cref{rtilde}.  Note that when an individual opinion changes from $x$ to $x'=x+c$, the opinion distribution changes by 
\begin{equation}
\delta f(z)=\delta(z-x-c)-\delta(z-x)
\label{indivdf}
\end{equation}
where $\delta(z)$ is the Dirac delta function.  Representation (\cref{genrep}) is then obtained in terms of functional derivatives of the election by substituting \cref{indivdf} into \cref{largen2}:
\begin{equation}
\label{rcd}
r_c(f,x)=\frac{1}{c}\delta y=\frac{1}{c}(\frac{\delta y}{\delta f(x+c)}-\frac{\delta y}{\delta f(x)})
\end{equation}
which, combined with \cref{limrep}, yields \cref{rtilde} (reproduced below).
\begin{equation}
r(f,x)=\frac{d}{dx}\frac{\delta y}{\delta f(x)}
\end{equation}
By the fundamental theorem of calculus, we see from \cref{rcd} that $r_c(f,x)$ is the average of $r(f,x)$ over $[x,x+c]$.

We now prove \cref{sumtoone} ($\int_{-\infty}^{\infty} f(x) r(f,x)dx=1$), which holds when the election ($y[f]$) is differentiable. 
For small $\epsilon$, $f(z+\epsilon)= f(z)+\epsilon\frac{d}{dz}f(z)+O(\epsilon^2)$, and thus, using \cref{largen2},
\begin{widetext}
\begin{equation}
y[f(z+\epsilon)]=y[f(z)+\epsilon\frac{d}{dz}f(z)+O(\epsilon^2)]=y[f(z)]+\epsilon\int_{-\infty}^{\infty}  \frac{df(x)}{dx}\frac{\delta y}{\delta f(x)}dx+O(\epsilon^2)
\end{equation} 
\end{widetext}
Since $f$ has compact support, which follows from $f$ being an approximation of the opinions of a finite number of citizens,
integrating by parts yields 
\begin{equation}
y[f(z+\epsilon)]=y[f(z)]-\epsilon\int_{-\infty}^{\infty} f(x) r(f,x)dx+O(\epsilon^2)
\end{equation}
which can be combined with \cref{invar} in the limit $\epsilon\rightarrow0$ to yield \cref{sumtoone}.  This proof assumes that $f$ is differentiable for illustrative purposes, but the result will more generally hold if $f$ and $r$ are treated as distributions (generalized functions).

\subsection{Nash equilibria of the electoral game}
\label{nash}
Consider a two-candidate election with endogenous candidacy: candidate positions (or, equivalently, candidates) are chosen in order to maximize the probability of victory. In this framework, the winner of the election will have adopted an unbeatable position $y^*$, provided such a position exists (i.e. a candidate with position $y^*$ will have at least a 50\% chance of winning against a candidate with any other position).  
Formally, $(y^*, y^*)$ is a \textit{Nash equilibrium}, since no candidate can improve her chances by changing her position.
Because the voting game is symmetric, if $(y_1,y_2)$ is a Nash equilibrium, then so are  $(y_1,y_1)$ and $(y_2,y_2)$; see, for instance, section~2.3 of~\cite{Coughlin1981}.  Thus, if there is a unique Nash equilibrium, it must be of the form $(y^*,y^*)$. 

\subsection{The utility difference model}
\label{udmexamples}
Building upon the two-candidate election framework above (\cref{nash}), we describe the assumptions behind the utility difference model given by \cref{ueqm} and give examples of its applications to median voting, mean voting, and an election between median and mean voting.  

Denoting the probability that a vote from someone with opinion $x$ will go to the first candidate minus the probability that it will go to the second by $p_x(y_1,y_2)$ (where $y_1$ and $y_2$ are the opinions of the first and second candidates, respectively), we assume there exists some function $u_x$ such that 
 \begin{equation}
 \label{pdif}
 p_x(y_1,y_2)=u_x(y_1)-u_x(y_2)
 \end{equation}
This assumption yields an affine linear \textit{utility difference model}~\cite{Banks2005}, the potential election outcomes for which are given by \cref{ueqm}.  Essentially, this model assumes that the position that maximizes a candidate's margin of victory does not depend on the position of the other candidate.  
We note that in order for \cref{pdif} to be able to be interpreted as a difference in probabilities, the functions $u_x(y)$ must satisfy 
\begin{equation}
\label{uconstraint}
\max_{y_1,y_2}|u_x(y_1)-u_x(y_2)|\leq1
\end{equation}
so that $|p_x(y_1,y_2)|\leq1$ always holds.

For $u_x(y)=-a^2(y-x)^2$ where $a$ is a positive constant, the mean opinion is selected, since for a random variable $X$, $\mathbb{E}[(X-\mu)^2]$ is minimized for $\mu=\mathbb{E}[X]$.  (Here, $y$ and the support of $f$ must be confined to an interval of length at most $a^{-1}$ in order to satisfy \cref{uconstraint}.)  For $u_x(y)=-a|y-x|$ (where again, $a$ is a positive constant, and $y$ and the support of $f$ must be confined to an interval of length at most $a^{-1}$), the median opinion is selected (since the median minimizes $\mathbb{E}[|X-m|]$),  although by a different mechanism than the deterministic voting assumptions of the Median Voter Theorem~\cite{Black1948}.  (The Median Voter Theorem states that when political opinions lie in one dimension and everyone votes for the candidate whose opinion is closest to his, the optimal position for a candidate to take is that of the median voter.)  Both of these functions can be viewed as limiting cases of the hyperbolic $u_x(y)=-\sqrt{a^2(y-x)^2+b^2}$, with $b<<a$  approximating median voting and $b>>a$ approximating mean voting.  Under mean voting, for a citizen opinion either to the right of both candidates or to the left of both candidates, the farther away this opinion is, the stronger the citizen's preference between the two candidates.   Under median voting, the strength of this citizen's preference for one candidate over the other is independent of how far the citizen's opinion is from both candidates.  For the intermediate case, the strength of this citizen's preference gets stronger up to a point and then levels off as the citizen's opinion moves farther away from both candidates.  However, in actual elections, citizens with opinions that are far from both candidates may be more likely to abstain from voting (or vote for a third-party candidate), which is why \cref{ualien} may be more realistic.

\subsection{Representation in the utility difference model}
\label{udmrep}
In this section, we calculate representation for the utility difference model given by \cref{ueqm}.  We derive the first part of \cref{ralien}, and we then calculate representation for the examples given in \cref{udmexamples}.  To do so, we must assume there is a single possible election outcome $y^*$ (see \cref{nash}).  Then, from \cref{ueqm}, 
\begin{equation}
\label{ystar1}
y^*=\argmax_{\tilde{y}} \int_{-\infty}^{\infty}  u_x(\tilde{y})f(x)dx
\end{equation}
which implies
\begin{equation}
\label{ystar}
0=\int_{-\infty}^{\infty}  u_x'(y^*)f(x)dx
\end{equation}
Note that \cref{ystar1} satisfies $y[f(x)]=y[\lambda f(x)]$ for any positive constant $\lambda$ (scale invariance), and let $\tilde{f}(x)=f(x)/N$ where $N$ is the size of the electorate, so that $\int_{-\infty}^{\infty} \tilde{f}(x)dx=1$.  Considering the change that arises from the addition of a single individual with opinion $x_0$ to the population, we define $\delta y^*$ by 
\begin{equation}
\label{y+edy}
y^*+\delta y^*=y[f(x)+\delta(x-x_0)]=y[\tilde{f}(x)+\epsilon \delta(x-x_0)]
\end{equation}
where $\epsilon=1/N$.
Thus, substituting $y^*+\delta y^*$ for $y^*$ and $\tilde{f}(x)+\epsilon \delta(x-x_0)$ for $f(x)$ into \cref{ystar}, 
\begin{equation}
\label{ystardystar}
0=\int_{-\infty}^{\infty}  u_x'(y^*+\delta y^*)(\tilde{f}(x)+\epsilon \delta(x-x_0))dx
\end{equation}
Because $\int_{-\infty}^{\infty}  u_x(x)f(x)dx$ is differentiable in $f$ and has a single maximum in $y$, expanding \cref{ystardystar} to lowest order in $\epsilon$ yields 
\begin{equation}
\label{dystar}
\delta y^*=\epsilon\frac{-u_{x_0}'(y^*)}{\int_{-\infty}^{\infty}  u_x''(y^*)\tilde{f}(x)dx}+O(\epsilon^2)
\end{equation}
Noting that the denominator is independent of $x_0$, of order $1$ (i.e. independent of $N$), and negative (otherwise, $y^*$ would be a minimum rather than a maximum),
\begin{equation}
\frac{\delta y}{\delta \tilde{f}(x_0)}=\lim_{\epsilon\rightarrow0}\frac{\delta y^*}{\epsilon} \propto u_x'(y^*)
\end{equation}
So, using \cref{rtilde},
\begin{equation}
r(f,x)=\frac{d}{dx}\frac{\delta y}{\delta f(x)}=\frac{1}{N}\frac{d}{dx}\frac{\delta y}{\delta \tilde{f}(x)}\propto \frac{1}{N}\frac{d}{dx}u_x'(y^*)
\end{equation}   
If $u_x(y)=u(y-x)$, as it must be for some function $u$ if the election is translationally invariant as in \cref{invar}, then
\begin{equation}
\label{urep}
r(f,x)\propto -\frac{1}{N}u''(y^*-x)
\end{equation}

Eq.~\ref{urep} provides a direct link between citizen preferences and the representation of opinions.  (If needed, the constant of proportionality can be determined through \cref{sumtoone}.)  Consider the examples for $u$ given in \cref{udmexamples}.  For $u(y-x)=-a^2(y-x)^2$ and $u(y-x)=-a|y-x|$, we can quickly derive the representation of opinions under mean voting ($r(f,x)=\frac{1}{N}$) and median voting ($r(f,x)=\frac{\delta (x-m)}{f(m)}$ where $m$ is the median of $f$), respectively.  For $u(y-x)=-\sqrt{a^2(y-x)^2+b^2}$, which yields an outcome between that of median and mean voting,
 \begin{equation}
 \label{intermediate}
 r(f,x)\propto (1+\frac{a^2}{b^2}(x-y[f])^2)^{-3/2}
\end{equation}
resulting in the representation of opinions being concentrated around the election outcome, but not infinitely concentrated as it is for median voting.  For $u$ that are not concave, there will exist some $x$ such that $u''(x-y)>0$, and representation will be negative for those opinions (see \cref{urep}).

\subsection{Instability in the utility difference model}
\label{instabproofs} 
Here, we explore the conditions under which instability can arise in the model described by \cref{ueqm}, and we elaborate on the concrete model of instability with outcomes given by  \cref{instaboutcome}.  For the model given by \cref{ueqm}, $r(f,x)=\frac{d}{dx}\frac{\delta y}{\delta f(x)}$ is shown to be well-defined as long as $y[f]$ is single-valued, i.e. \cref{ueqm} has a single maximum (\cref{udmrep}).  Thus, instability can occur only when there are multiple maxima.
(For a single maximum $y^*$ with $\int_{-\infty}^{\infty}  u''(x-y^*)f(x)dx=0$, the functional derivative of $y$ is not defined, but, as can be shown in a higher order analysis, there is no instability.  In particular, for $\delta y^*$ defined by \cref{y+edy}, we can derive $$\frac{1}{6}(\delta y^*)^3=\frac{-\epsilon u_{x_0}'(y^*)}{\int_{-\infty}^{\infty}  \tilde{f}(x) u_x''''(y^*)dx}+O(\epsilon^{4/3})$$ in place of \cref{dystar}, which yields $\delta y^*\propto\epsilon^{1/3}u_{x_0}'(y^*)^{1/3}+O(\epsilon^{2/3})$.  Note that $$r(f,x_0)=\frac{d}{dx}\Bigr|_{x=x_0}\delta y^*$$ is well-defined for any given $\epsilon=1/N$---although it is not given by \cref{rtilde} since $\frac{\delta y}{\delta f(x)}$ does not exist---thus, there is no instability.)

For an opinion distribution in which \cref{ueqm} has two maxima $y_1^*$ and $y_2^*$, if each position is taken by one candidate, then each candidate has a 50\% chance of winning the election.  However, an arbitrarily small change in the opinion distribution can favor one outcome over the other, giving either $y_1^*$ or $y_2^*$ a chance of winning that is arbitrarily close to 100\% in the large-population limit.  (For a finite population, the outcome of the election in this model is not deterministic, and the probability of a given candidate winning is continuous with respect to the opinion distribution.  This discontinuity in $y[f]$ arising from multiple Nash equilibria in the large-population limit is analogous to a first-order phase transition.)

The existence of multiple maxima in $y$ implies that $\int_{-\infty}^{\infty}  u(x-y)f(x)dx$ is not concave (ignoring the degenerate case in which $\int_{-\infty}^{\infty}  u(x-y)f(x)dx$ is constant over some interval).  Thus, instability can arise only in the case of non-concave $u$, which is precisely the same condition under which negative representation occurs.  That instability can arise only in the presence of negative representation should not surprise us, since it was proven under more general conditions in the main text.  For this class of models, we also find that negative representation implies that there exist distributions of opinions for which instability arises, i.e. if $u$ is not concave, then there exists an $f(x)$ such that \cref{ueqm} has multiple maxima.  To see why this is true, consider an opinion distribution $f$ such that $f(x)=f(-x)$.  Then, if there is a single maximum of \cref{ueqm}, it must lie at $y^*=0$.  In order for $y^*=0$ to be a maximum, we must have $\int_{-\infty}^\infty f(x)u''(x)dx=2\int_0^\infty f(x)u''(x)dx\leq0$ (where the equality follows from the symmetry of $u$ and $f$).  But, assuming $u$ is twice continuously differentiable and not concave, there exists an $f$ such that $f(x)=f(-x)$ and $\int_0^\infty f(x)u''(x)dx>0$.  For such an $f$, $y^*=0$ is not a maximum, thus contradicting our assumption that there was a unique maximum.

To provide an example of how, for non-concave $u$, the election is unstable for certain opinion distributions, we consider the $u$ used for the example of negative representation: $u(y-x)=\exp[-\frac{(y-x)^2}{2a^2}]$ for some positive constant $a$ (\cref{ualien}).
We take the distribution of electorate opinions to be a sum of two (potentially unequally weighted) normal distributions of equal variance:
\begin{equation}
\label{sbimodalf}
f(x)=\sum_{\alpha=1,2} w_\alpha e^{-\frac{(x-\mu_\alpha)^2}{2\sigma^2}}
\end{equation}
Then, from \cref{ueqm}, the outcome of the election is then given by 
\begin{equation}
\label{argmax}
\argmax_y\int_{-\infty}^{\infty}  f(x) e^{-\frac{(y-x)^2}{2a^2}}dx=\argmax_y\sum_{\alpha=1,2} w_\alpha e^{-\frac{(y-\mu_\alpha)^2}{2(a^2+\sigma^2)}}
\end{equation}
Without loss of generality, we can assume that $\mu_2=-\mu_1\equiv \Delta\geq 0$.  Defining the normalized election outcome  $\hat{y}\equiv y/\Delta$, we solve \cref{argmax} to get the following condition:
\begin{equation}
\label{instabyhat}
\hat{y}=\tanh(J\hat{y}+h)
\end{equation}
where $J=\Delta^2/(a^2+\sigma^2)$ and $h=\frac{1}{2}\ln\frac{w_2}{w_1}$.  For $w_1=w_2$ (i.e. for equally sized subpopulations), $h=0$, and the election is stable with $\hat{y}=0$ for $J\leq1$ and unstable for $J>1$.  In the unstable regime, there are two possible outcomes described by $\pm|\hat{y}^*|$, and an arbitrarily small change in $h$ can cause $\hat{y}$ to swing between its positive and negative value.  In this regime, the majority of one of the subpopulations will be negatively represented: for $y^*<0$, over half of the subpopulation centered at $\Delta$ will have opinions $x$ with $x-y^*>\Delta-y^*>a$ ($\Delta>a$ in the unstable regime), and, from \cref{ralien}, representation is negative for these opinions.  Likewise, for $y^*>0$ in the unstable regime, over half of the subpopulation centered at $-\Delta$ will be negatively represented.

\subsection{Connection with the mean-field Ising model}
\label{ising}
We map the voting model that gives rise to \cref{instabyhat} (\cref{instaboutcome} in the main text) onto a mean-field Ising model~\cite{Kadanoff2009}.  To begin constructing this map, note that the left-hand side of \cref{argmax} gives the limiting value of $y$ as $N\rightarrow\infty$ when $y$ is drawn from a probability distribution corresponding to the partition function
\begin{equation}
Z=\int_{-\infty}^{\infty}  \qty[\int_{-\infty}^{\infty}  e^{-\frac{(y-x)^2}{2a^2}} f(x)dx]^Ndy
\end{equation}

Because $y$ is a gaussian random variable, we can explicitly integrate over $y$, which yields, up to a multiplicative constant,
\begin{widetext}
\begin{equation}
\label{eq:mfspins}
Z=\int_{-\infty}^{\infty} e^{-\sum_i\frac{(x_i-\bar{x})^2}{2a^2}}  \Pi_{i=1}^N(f(x_i)dx_i)=\int_{-\infty}^{\infty}  e^{-\frac{1}{2N}\sum_{i,j}\frac{(x_i-x_j)^2}{2a^2}}\Pi_{i=1}^N(f(x_i)dx_i)
\end{equation}
\end{widetext}
This equation describes $N$ interacting probabilistic ``spins,''  with each spin weighted by the opinion distribution $f(x)$, with an energy penalty proportional to its mean-square distance from all of the other spins.  For $f(x)$ given by \cref{sbimodalf} (\cref{sumgauss} in the main text), the behavior is exactly that of a mean-field Ising model (with an external magnetic field for $w_1\neq w_2$);
in general, for bimodal symmetric $f(x)$ we expect a phase transition in which the system will spontaneously break the symmetry between the peaks as the peaks move farther apart.  In the stable/disordered phase, both of the peaks of $f(x)$ are sampled by the ``spins;'' in the unstable/ordered phase, however, only one of the two peaks is sampled, and therefore the other peak is not represented.  Despite the fact that each individual votes independently from everyone else, citizens are coupled through their collectively choosing a candidate, which is reflected by the effective interactions between the ``spins.'' In the limit of weak interactions, i.e. $a\rightarrow \infty$, we recover mean voting, since in this limit, $u(y-x)=\exp[-\frac{(y-x)^2}{2a^2}]\approx 1-\frac{1}{2a^2}(y-x)^2$.  (Quadratic utility functions yield mean voting---see \cref{udmexamples}.)  Thus, mean voting is a way of ``independently'' aggregating opinions. 

For a general $u(y-x_i)=\exp[-V(y-x_i)]$ (we can always write $u$ in this form with $V(x)\geq c$ for some $c\in\mathbb{R}$), we note that \cref{ueqm} yields an election outcome equivalent to the limit of $y$ as $N\rightarrow \infty$ with $y$ drawn from 
\begin{widetext}
\begin{equation}
\label{eq:mfspinsV}
Z=\int_{-\infty}^{\infty} \qty[\int_{-\infty}^{\infty} e^{-V(y-x)}f(x)dx]^N dy=\int_{-\infty}^{\infty}  e^{-\sum_iV(y-x_i)} \Pi_{i=1}^N(f(x_i)dx_i)
\end{equation}
\end{widetext}
For quadratic $V$, we saw above that we could exactly integrate over the election outcome $y$ to yield pairwise quadratic interactions between the $x_i$ variables, but for general $V$, such an integration will yield many interaction terms of higher than quadratic order between these $x_i$.  Although such integration cannot be carried out precisely, we expect this interacting system to undergo a phase transition for bimodal $f(x)$ if the expansion of $V$ produces sufficiently strong interactions.  Thus, the system's behavior should be similar to the exactly solvable case in which $V$ is quadratic.

\subsection{Empirical data}
\label{empirics}
To determine if the stability of U.S. presidential elections has changed over time, we used data from Jordan \etal~\cite{Jordan2014} on the polarization in the party platforms.  Jordan \etal~used a combination of machine learning and human judgment to determine which of the frequently used words in the party platforms were polarizing and then determined the number of polarizing words (classified by political issue dimensions such as economic, foreign, etc.) in the Republican and Democratic platforms from 1944 to 2012.  From this data, we calculated the total number of polarizing words as a percentage of all words in the platforms.  We chose the percentage of polarizing words in the party platforms as a measure of political polarization over other measures of ideology---such as NOMINATE scores~\cite{Poole1985}, which measure ideological purity based on agreement with other politicians---because we wanted an external, content-based measure of divergence in opinion rather than a measure of ideology that depends only on the positions that politicians take relative to one another.   
To construct \cref{empirical}, we plotted by year the fraction of polarizing words in the Democratic platforms and the negative of the fraction of polarizing words in the Republican platforms.  To correct for any time-independent bias affecting the number of polarizing words in the party platforms---for instance, which words Jordan \etal~designated as polarizing---we subtracted from the data for each party separately the fraction of polarizing words from the year of least polarization for that party (which was 1948 for both parties).  Thus, the data shown are the changes in the fraction of polarizing words relative to their baseline value (0.0258 for Democratic platforms and 0.0693 for Republican platforms).

As was noted in the main text and explained in \cref{ising}, our voting model (\cref{bifurcation}) is equivalent to a mean-field Ising model.  Real-world elections are unlikely to follow this model exactly, and even if they did, there is no reason to believe that there would be a simple relationship between polarization (for which $J$ is a dimensionless measure) and time.  Nonetheless, if U.S. presidential elections underwent a phase transition in the same \textit{universality class} as the mean-field Ising model, then in the vicinity of the phase transition, polarization would increase in proportion to the square root of the time from the transition, regardless of the precise relationship between time and polarization.   In much the same way, magnetization increases near a ferromagnetic phase transition in proportion to $(T-T_c)^\beta$, where $T$ is temperature, $T_c$ is the temperature at which the phase transition occurs, and $\beta$ is known as a \textit{critical exponent}, which depends only on the universality class to which the phase transition belongs~\cite{Kardar2007}.  Inspired by this universality, we fit the polarization of both parties to the piecewise function 
\begin{equation}
f(x)= \begin{cases} 
      0 & x\leq x_0 \\
      A\sqrt{x-x_0} & x>x_0
   \end{cases}
\end{equation}
where $x$ is the year and $f(x)$ is the fraction of polarizing words in that year's platform relative to the baseline value (see above), and $A$ and $x_0$ are free parameters, corresponding to the amplitude of the polarization and the year that it begins, respectively.  We found that $x_0=1970.54$ and $A=0.0079196$ minimize the total sum of square errors, yielding $R^2$ values of $0.86$ for the Democratic party and $0.89$ for the Republican party.  If the two parties are considered together, $R^2=0.87$.

\vspace{2em}
\section{Supplementary Text}

\subsection{Multidimensional opinion space}
\label{multi}
For the sake of simplicity, this paper focuses on systems with a one-dimensional opinion space, but the concepts developed in this paper can naturally be extended to a multidimensional opinion space, where the opinions of the electorate and candidates lie in $\mathbb{R}^n$, as in~\cite{Coughlin1981, Enelow1989, Banks2005,Enelow1984}.  This extension will be briefly outlined here.  The definition of representation is generalized by replacing \cref{genrep} with
\begin{equation}
    r_{\vec{c},\mu\nu}(f,x)=\frac{\delta y_\mu c_\nu}{|c|^2}
\end{equation}
For a scalar measure, we use
\begin{equation}
\label{tracerep}
\tr[r_{\vec{c}}]=\frac{\vec{c}\cdot\delta \vec{y}}{|c|^2}
\end{equation}
When there exists an $r_{\mu\nu}(f,x)$ such that
\begin{equation}
\label{multilimrep}
\delta y_\mu=\sum_\nu r_{\mu\nu}(f,x)c_\nu+O(c^2)
\end{equation}
for all $\vec{c}$, this $r_{\mu\nu}(f,x)$ can be used in place of \cref{limrep} as a representation independent of $\vec{c}$.
In the large-population limit, \cref{rcd} is then replaced (using Einstein-summation notation) by the path-independent integral 
\begin{widetext}
\begin{equation}
\tr[r_{\vec{c}}(f,x)]=\frac{c_\mu}{|c|^2}\qty(\frac{\delta y_\mu}{\delta f(\vec{x}+\vec{c})}-\frac{\delta y_\mu}{\delta f(\vec{x})})=\frac{c_\mu}{|c|^2} \int_{\vec{x}}^{\vec{x}+\vec{c}}r_{\mu\nu}(f,x')dx_\nu'
\end{equation}
\end{widetext}
where $r(f,x)$ (which satisfies \cref{multilimrep}) is a matrix defined by
\begin{equation}
\label{multidrep}
r_{\mu\nu}(f,x)=\frac{\partial}{\partial x_\nu}\frac{\delta y_\mu}{\delta f(x)}
\end{equation}
The differential representation in a direction given by the unit vector $\hat{v}$ is then given by $\hat{v}_\mu r_{\mu\nu}(f,x)\hat{v}_\nu$, which yields the same results as eq. 7 of~\cite{Napel2004} in the limit of a continuum of voters.  The trace $\tr[r]$ gives a rotationally invariant scalar measure.

The representation normalization condition corresponding to \cref{sumtoone} is $\int  f(x) r_{\mu\nu}(f,x) dx=\delta_{\mu\nu}$ where the integral is taken over $\mathbb{R}^n$.

In the multidimensional case, instability also implies a failure in representation.  In a manner analogous to \cref{instab}, instability is characterized by 
\begin{equation}
\lim_{\vec{c}\rightarrow 0} c_\nu r_{\vec{c},\mu\nu}\neq 0
\end{equation} 
Generally, instability implies either that $\lim_{\vec{c}\rightarrow 0} |c|\tr[r_{\vec{c}}]\neq 0$, in which case negative representation (defined by $\tr[r_{\vec{c}}]<0$) follows in the same way as the one-dimensional case, or that an infinitesimal change in opinion causes a finite orthogonal change in the outcome of the election. 
In this case, by considering further infinitesimal changes in opinion parallel to the first change in election outcome, and assuming that the magnitude of the change in the outcome of the election cannot grow without bound, one either gets negative representation directly ($\tr[r_{\vec{c}}]<0$ for some $\vec{c}$)---or $\tr[r_{\vec{c}}]>1$, from which negative representation follows as it does in the one-dimensional case.

Just as in the one-dimensional case, increasing polarization can drive the election through a phase transition from a stable to unstable regime, as \cref{eq:mfspins,eq:mfspinsV} apply equally well if the opinions $x_i$ are vector quantities. The nature of the unstable regime will depend on the symmetries present in the problem.\footnote{However, as all of the possible phase transitions are described by a mean-field theory, they all belong to the same universality class.}  For example, consider a distribution of opinions given by
\begin{equation}
f(\vec{x})=\sum_{\alpha=1}^{N_\alpha}e^{-\frac{(\vec{x}-\vec{\mu}_\alpha)^2}{2\sigma^2}}
\end{equation}
where $N_\alpha$ is the number of subpopulations.  For $N_\alpha=2$, the instability that occurs for sufficiently large $|\vec{\mu}_1-\vec{\mu}_2|$ will be very similar to what was described in the main text for the one-dimensional case.  For $N_\alpha >2$, if the set of points $\{\vec{\mu}_\alpha\}$ possess permutation symmetry, the system will resemble that of a mean-field Potts model (assuming the $V$ in \cref{eq:mfspinsV} is isotropic).  However, unlike the case for $N_\alpha=2$---where regardless of the values of the $\vec{\mu}_\alpha$, the system possesses the appropriate symmetry to belong to the Ising universality class (due to the overall translational invariance of the problem and thus the ability to redefine the origin)---for $N_\alpha>2$, we should not expect the system to generally possess the permutation symmetry necessary for Potts universality.   Thus, we expect the real-world phase transitions that we observe to be of the $N_\alpha=2$ type, i.e. of the same type of instability observed for a one-dimensional opinion space.  For $N_\alpha>2$, the situation may be able to be approximated by $N_\alpha=2$, especially if there are dynamical social or political forces that pull the distribution of opinions towards a situation in which competition is roughly balanced between two opposing (potentially multi-party) coalitions at any given point in time.

\subsection{The Owen-Shapley index as a special case}
\label{consistency} 
In this section we show that there exist functionals $y[f(x)]$ for which our representation measure (\cref{rtilde}) reproduces the values of both the deterministic and probabilistic Owen-Shapley voting power indices.  Thus, these voting power indices can be thought of as special cases of our measure.  We give a brief background on the voting power literature and then consider the case of a one-dimensional opinion space, followed by a generalization to the case of a multidimensional opinion space for which the Owen-Shapley index was primarily designed.

When nothing is known about the preferences of voters, their political power has traditionally been measured by \textit{a priori voting power}~\cite{Felsenthal2004}, which reflects the probability that a given individual or entity will cast the deciding vote and is usually measured by   
either the \textit{Penrose index}~\cite{Penrose1946} or the \textit{Shapley-Shubik index}~\cite{Shapley1954}.  More precisely, to calculate a voter's \textit{a priori} voting power, consider a random division of the rest of the voters into two camps.  Then the probability that the excluded voter will get his way regardless of which camp he joins is his voting power;  the Penrose index and the Shapley-Shubik index differ only in the way in which they randomly choose a division.  While the Penrose index assumes that each voter randomly chooses one side or the other, the Shapley-Shubik index re-weights the probabilities so that each ordering of voters is equally likely.\footnote
{There are some fundamental differences in the motivation behind the indices~\cite{Felsenthal2004}, but mathematically, they are rather similar, though neither is without drawbacks:  while the assumptions behind the Penrose index are simpler, in general the sum of all voters' Penrose indices will not equal $1$, while the sum of all voters' Shapley-Shubik indices will.}  
These indices provide useful and counterintuitive results when the voters possess differing numbers of votes, as in the European Union.  For instance, under the 1958 voting rules for the European Economic Community, Luxembourg, despite having had one vote, had no voting power, since there were no possible divisions of the other five countries such that Luxembourg's vote would be decisive~\cite{Gelman2002}.  But in elections in which each voter has one vote, all measures of \textit{a priori} voting power result in each voter having an equal amount of power.  A measure of voting power that takes voter preferences into account is needed to determine how various opinions are differently represented.   Many preference-based measures have been proposed---for instance, the spatial Shapley-Shubik index, also known as the Owen-Shapley index~\cite{Owen1989}---but, as we will see, these measures implicitly assume that people vote in a particular way.  

In one dimension, the deterministic Owen-Shapley index~\cite{Owen1989} allows for only two possible orderings of the voters (left to right or right to left), for which the median voter is pivotal in both, thus yielding the same concentration of power that our representation measure (\cref{limrep}) yields in the case of median voting ($y[f]=\median[f]\equiv m$ yields $r(f,x)=\frac{d}{dx}\frac{\delta y}{\delta f(x)}=\frac{\delta (x-m)}{f(m)}$).   Benati and Marzetti~\cite{Benati2013} note that this extreme concentration of power is due to the deterministic nature of the Owen-Shapley model, which assigns zero probability to almost all orderings.  They propose a generalized election model in which voters' opinions have both a deterministic and a random component.  In the one-dimensional case, their treatment is equivalent to denoting the probabilistic opinion $X_i$ of voter $i$ by 
\begin{equation}
\label{probopinion}
X_i=x_i+\epsilon_i
\end{equation}
where the $x_i$ are deterministic and the $\epsilon_i$ are independent random variables with a continuous probability density function $f_\epsilon(\epsilon_i)$.  Denoting the distribution of the $x_i$ over the population by $f(x_i)$ (note that $f$ will be a sum of delta functions for a finite population) and choosing an election in which people vote for the candidate closest to their probabilistic opinions $X_i$,\footnote
{This election takes the form of the \textit{Random Utility Model} mentioned in section~2.2 of~\cite{Banks2005}.}
the Nash equilibrium for the two candidates' opinions is the median of the distribution $f_X(X)\equiv(f\ast f_\epsilon)(x)=\int_{-\infty}^{\infty} f(x)f_\epsilon(X-x)dx$, i.e.
\begin{equation}
y[f]=\median[f_X]=\median[f\ast f_\epsilon]\equiv m
\end{equation}
The continuity of $f_X$ follows from that of $f_\epsilon$, and we have made the additional assumption that $f_X(m)\neq0$; otherwise, there is no unique Nash equilibrium.

We then calculate 
\begin{widetext}
\begin{equation}
\frac{\delta y[f]}{\delta f(x)}=\int_{-\infty}^{\infty}  \frac{\delta\median[f_X]}{\delta f_X(z)} \frac{\delta f_X(z)}{\delta f(x)}dz=\int_{-\infty}^{\infty} \frac{\sign(z-m)}{2f_X(m)} f_\epsilon(z-x)dz=\int_{-\infty}^{\infty}  \frac{\sign(z+x-m)}{2f_X(m)}f_\epsilon(z)dz
\end{equation}
which yields the representation measure (\cref{rtilde}) for opinion $i$:
\begin{equation}
\label{probmed}
r(f,x_i)=\frac{d}{dx_i}\frac{\delta y[f]}{\delta f(x_i)}=\int_{-\infty}^{\infty} \frac{2\delta(z+x_i-m)}{2f_X(m)} f_\epsilon(z) dz=\frac{1}{f_X(m)}f_\epsilon(m-x_i)
\end{equation}
\end{widetext}
The rightmost side of \cref{probmed} is the probability that voter $i$ is the median---i.e. pivotal---voter; thus, $r(f,x_i)$ is equal to the generalized Owen-Shapley index for voter $i$. 

The Owen-Shapley index was developed primarily for multidimensional opinion spaces in $\mathbb{R}^n$.  Owen and Shapley~\cite{Owen1989} consider a randomly drawn unit vector $\hat{v}\in\mathbb{R}^n$ and then order individuals by defining $i<j$ if $\vec{X_i}\cdot\hat{v}<\vec{X_j}\cdot \hat{v}$.  (The $\vec{X_i}$ are deterministic but can easily be modified to be partially probabilistic as in~\cite{Benati2013}.)  The Owen-Shapley index of $i$ is again the probability that $i$ is the median of the resulting ordering.  To see how this power index is a special case of our multidimensional representation measure (\cref{multidrep}), consider the following method of choosing a candidate, given a set of voters with (potentially probabilistic) opinions $\vec{X_i}\in\mathbb{R}^n$:

1) Randomly choose an orthonormal basis $V=\{\hat{v}_1,...,\hat{v}_n\}$ for $\mathbb{R}^n$.

2) Conditioning on the orthonormal basis $V$, let $m_\alpha(V)$ be the median of the probability distribution function for $\hat{v}_\alpha\cdot \vec{X}$, where $\vec{X}$ is randomly drawn from the voter opinions $\vec{X_i}$.

3) The election outcome is then given by $\vec{y}[f]=\sum_{\alpha=1}^n m_\alpha(V)\hat{v}_\alpha$.

Note that $\vec{y}$ is now a random variable (since it depends on the orthonormal basis $V$), and so the right-hand side of \cref{multidrep} must be replaced by its expectation value, i.e. $r_{\mu\nu}(f,\vec{x})=\mathbb{E}[\frac{\partial}{\partial x_\nu}\frac{\delta y_\mu}{\delta f(\vec{x})}]$.  From \cref{probmed} (with $\hat{v}\cdot\vec{y}$, $\hat{v}\cdot\vec{x_i}$, and $\hat{v}\cdot{\vec{\epsilon_i}}$ substituting for $y$, $x_i$, and $\epsilon_i$), $\hat{v}_\mu r_{\mu\nu}(f,\vec{x}_i)\hat{v}_\nu$ is equal to the probability that $i$ will be the median voter along $\hat{v}$.  Therefore, $\tr[r(f,\vec{x}_i)]$ is equal to the expected number of basis vectors along which $i$ will be the median, and so $\tr[r(f,\vec{x}_i)]$ is equivalent to $n$ times the Owen-Shapley index.
 
The agreement between $r(f,\vec{x})$ and the Owen-Shapley index follows from the fact that $\frac{d}{d(\hat{v}\cdot\vec{x}_i)}\frac{\delta \median[f_{\hat{v}\cdot\vec{X}}]}{\delta f(\vec{x}_i)}$ measures the probability that $i$ is the median voter along $\hat{v}$ for this class of voting models.  In this sense, the Owen-Shapley index of power implicitly assumes an election in which some sort of median is chosen.  This model is appropriate when voters vote deterministically (although the options presented for them to vote on may be random).  But such deterministic voting assumes that voters distinguish between very small differences in policy with 100\% certainty, and it also assumes that there is no chance that a voter abstains.  While these assumptions may hold for assemblies of elected officials (and in particular to the EU, where these measures are most commonly applied), they tend to fail for mass elections, in which a citizen may sometimes vote for the candidate farther from her opinion and sometimes may choose not to vote at all. 

\subsection{A brief review of social choice theory}
\label{screview}
Here, we briefly review the social choice literature on elections, focusing in particular on the aspects of spatial models of elections---i.e. models that denote the political preferences of a citizen by a point in some space that can be embedded in $\mathbb{R}^n$ for some $n$---that are most relevant to our work.  (These points are often referred to as \textit{ideal points}---rather than \textit{opinions}, as in this manuscript---in order to emphasize the simplification that takes place when opinions are placed in a low-dimensional space.)  The social choice literature also considers more general questions concerning the aggregation of opinions, often with counterintuitive results~\cite{de2014essai,arrow2012social,gibbard1973manipulation,satterthwaite1975strategy,elster1989foundations};\footnote{For example, Arrow's Impossibility Theorem~\cite{arrow1950} states, roughly speaking, that a dictatorship is the only system of aggregating preferences that 1) prefers A to B if the citizens unanimously prefer A to B, and 2) does not allow its preference between two alternatives A and B to be affected by citizens preferences concerning a third alternative C.}
see~\cite{sep-social-choice} for an overview. The simplest spatial models of elections take $n=1$, i.e. they assume that political preferences can be determined by a citizen's position on a line (often interpreted as the left-right political spectrum), e.g.~\cite{Hotelling1929,Black1948,Downs1957}.  These models assume that there are two candidates competing for a majority of the votes and that all citizens vote and that they vote for whichever candidate is closest to them.  Under these conditions, there is a pure Nash equilibrium (see Methods section `Nash equilibria of the electoral game') in the candidate strategies, which is for both candidates to adopt the position of the median voter.  Downs~\cite{Downs1957} recognized the limitations of these assumptions and intuited qualitatively that due to the nonvoting that can arise from alienation, ``democracy does not lead to effective, stable government when the electorate is polarized.''

Subsequent work has focused on the existence of Nash equilibria in candidate strategies under various assumptions concerning voter and candidate motivations.  In a multi-dimensional opinion space ($n>1$), the deterministic voting that gives rise to the Median Voter Theorem for $n=1$ does not yield a Nash equilibrium without additional restrictions on the voters' preferences (see e.g. ~\cite{plott1967,davis1970,davis1972social}).  Nash equilibria can often be restored by considering spatial models in which citizens cast votes probabilistically, since probabilistic voting allows for smoother changes in voting behavior with respect to changes in the candidate positions (e.g.~\cite{intriligator1973,Coughlin1981,Enelow1989,lin1999,Coughlin1992}; see~\cite{Banks2005} for a review).  Other assumptions include, for instance, voter abstention~\cite{hinich1972,hinich1973}, policy-motivated candidates (e.g.~\cite{calvert1985robustness,lindbeck1993model}) and game-theoretic considerations on the part of the voters (e.g.~\cite{ledyard1984,mckelvey2006theory}).

In our work, we are concerned with the the functional $y[f(x)]$, which maps the distribution of citizen opinions onto the electoral outcome, rather than with the mechanism that gives rise to this map.   In the latter half of the main text, we build upon the probabilistic voting models developed in the literature, relaxing the assumption of concave voter preferences so as to provide a concrete demonstration of how negative representation and electoral instability can arise, the latter of which is reflected in the existence of \textit{multiple} Nash equilibria.

\bibliography{refs}
\bibliographystyle{unsrt}

\end{document}